\documentstyle[a4]{article}

\input epsf
\input amssymb.sty

\def\al{\alpha}

\def\om{\omega}

\def\map{\rightarrow}
\def\bq{\begin{equation}}
\def\eq{\end{equation}}
\def\li{\langle}
\def\ri{\rangle}

\def\ti{\tilde}
\def\Om{\Omega}
\def\Si{\Sigma}
\def\La{\Lambda}
\def\ss{\subset}
\def\ovl{\overline}
\def\Hi{{\cal H}}

\begin{document}

\begin{titlepage}
%\flushright{CERN-TH/2000-240}
{}~\vskip 1.5cm
\begin{center}
{\LARGE (Non-)Abelian Kramers--Wannier Duality\\ And Topological Field Theory}
\vskip 2cm
{\large Pavol \v Severa\footnote{email: pavol.severa@cern.ch \\ The author is supported by the European Postdoctoral Institute for Mathematical Sciences}\\}
\vskip 0.5cm
{\small \it CERN - TH Division\\ CH-1211 Geneva 23\\ Switzerland}
\vskip 3cm
\end{center}
\begin{abstract}
We study a connection between duality and topological field theories. First, 2d Kramers--Wannier 
duality is formulated as a simple 3d topological claim (more or less Poincar\'e duality), and a similar 
formulation is given for higher-dimensional cases. In this form they lead to 
simple TFTs with boundary coloured in two colours. Classical models (Poisson--Lie T-duality) suggest a non-abelian generalization in the 2d 
case, with abelian groups replaced by quantum groups. Amazingly, the TFT 
formulation solves the problem without computation: quantum 
groups appear in pictures, independently of the classical motivation. Connection with Chern--Simons theory appears at the symplectic level, and also in the pictures of 
the Drinfeld double: Reshetikhin--Turaev invariants of links in 3-manifolds, computed from the double, are included in these TFTs. All this suggests nice phenomena in higher dimensions.
\end{abstract}
\end{titlepage}

\section{Introduction: KW duality as a 3d topological claim}
Kramers--Wannier (KW) duality in 2d statistical models can be formulated as a simple topological claim about 
pictures like this one:

$$\epsfxsize 6cm \epsfbox{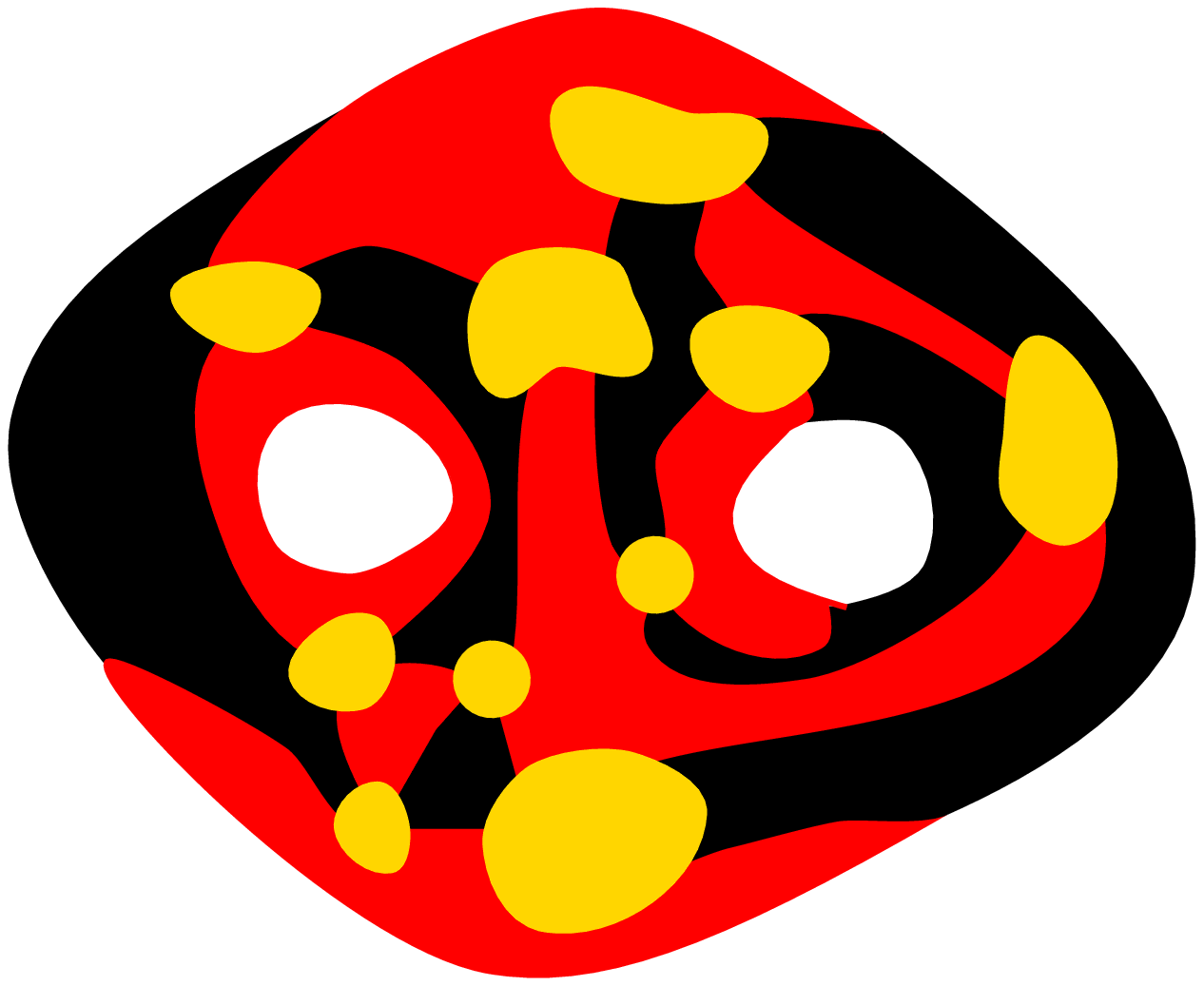}$$
The picture represents a 3d body (a ritual mask) made of yellow material, with the surface partially painted in  red and black. For definiteness imagine that the invisible side is unpainted (i.e. completely yellow). In general we have a compact oriented yellow 3-fold $\Om$ with the 
boundary coloured in this way (in a locally nice way: the borders of 
the colour stains are piecewise linear (say) and at most three of them meet at 
a single point).  

We choose a finite abelian group $G$ and its dual $\ti G$. Let $y$ be the yellow 
 part of the boundary; it is an oriented surface with the boundary 
coloured in black and red. The relative cohomology groups $H^1(y,r;G)$ and 
$H^1(y,b;\ti G)$ are mutually dual via Poincar\'e duality (in expressions like 
$H^k(X,r;G)$, $r$ denotes the red part of $X$, and $b$ the black part). Let 
$\rho:H^1(\Om,r;G)\map H^1(y,r;G)$ and $\ti\rho:H^1(\Om,b;\ti G)\map 
H^1(y,b;\ti G)$ be the restriction maps. According to KW duality, their images are 
each other's annihilators. It is an immediate consequence of Poincar\'e duality 
and of the exactness of
$$H^1(\Om,r;G)\map H^1(y\cup r,r;G)\map H^2(\Om,y\cup r;G).$$

In statistical models it is used in the following form: we take a function 
$f$ on $H^1(y,r;G)$ (the Boltzmann weight) and compute the partition sum
\bq Z(f)=\sum_{x\in H^1(\Om,r;G)}f(\rho(x)).\eq
Let $\hat f$ denote the Fourier 
transform of $f$. We can compute \bq\ti Z(\hat f)=\sum_{x\in H^1(\Om,b;\ti 
G)}\hat f(\ti\rho(x)).\eq KW duality says (via Poisson summation formula) that 
up to an inessential factor we have $Z(f)=\ti Z(\hat f)$.

Let us stop to make a connection with more usual formulations. Notice that an element 
of $H^1(X,Y;G)$ is the same as (the isomorphism class of) a principal $G$-bundle 
over $X$ with a given section over $Y\ss X$. If $\Om$ is a 3d ball (with 
coloured surface), an element of $H^1(\Om,r;G)$ is therefore specified by 
choosing an element of $G$ for each red stain. We may imagine that there is a 
$G$-valued spin sitting at each such  stain and, to compute (1), we take the sum 
over all their values (we overcount $|G|$ times, but it is inessential). According to KW 
duality, the same result can be obtained by summing over $\ti 
G$-spins at the black stains. The spins at red (or black) stains interact through the yellow stains. If all the yellow stains are as those visible on the picture (disks with two red and two black neighbours), we have the usual two-point interactions; for disks with more neighbours we would have more-point interactions.

Finally, let us look at the picture again. It does not represent a ball and the 
back yellow stain is not a disk. The Boltzmann weight for the back stain can be 
understood as the specification of the boundary and periodicity conditions on 
the visible surface (the $G$-bundle type together with sections over the red parts of the boundary); again there 
are spins at the red stains but the neighbours of the back yellow stain are not summed 
over -- they form the boundary condition.

These examples are more or less all that we would like; the general case seems 
to be general beyond usual applications. But it will come handy when we consider non-abelian generalizations.

The KW duality described up to now is only the $(1,1)$-version. For the
$(k,l)$-version (for statistical models in $k+l$ dimensions) we consider yellow $(k+l+1)$-dimensional $\Om$'s with $\partial \Om$ in the 
three colours as before (up to now only the combination $k+l$ enters). Instead 
of $H^1(\Om,r;G)$ and $H^1(\Om,b;\ti G)$ we take $H^k(\Om,r;G)$ and 
$H^l(\Om,b;\ti G)$. The claim and the proof of $(k,l)$-duality are as in the 
$(1,1)$-case.

What are we going to do? First of all,  expression (1) has the form of a very 
simple topological field theory (with boundary coloured in red and black), described in the next section. We shall then
look at the non-abelian version. In the $(1,1)$-case classical models suggest 
that the pair $G$, $\ti G$ should be replaced by a pair of mutually dual quantum 
groups. So we are faced with the difficult and somewhat arbitrary task of defining 
and understanding quantum analogues of cohomology groups and of the Poisson summation 
formula. But miraculously, none of these has to be done. Pictures alone (in the form of TFTs)  solve 
the problem and quantum groups appear. This suggests, of course, that this point 
of view might be interesting in higher dimensions, the $(2,2)$ case -- the 
electric--magnetic duality -- being of particular interest.

\section{KW TFTs and the squeezing property}
As we mentioned,  expression (1) (and its generalization to $(k,l)$) has the 
form of a TFT with boundary coloured in red and black. We understand TFT as defined by Atiyah \cite{Ati} and for definiteness we choose its hermitian version; nothing like central extensions is 
taken into account. To each oriented yellow $(k+l)$-dim $\Si$ with black-and-red 
boundary, we associate a non-zero finite-dimensional Hilbert space $\Hi(\Si)=L^2(H^k(\Si,r;G))$. And for 
each $\Om$ we have a linear form on the Hilbert space corresponding to $y$  -- 
the one given by (1). However, the normalization has to be changed slightly for 
the gluing property to hold (this is only a technical problem): we set
\bq Z_\Om(f)={1\over\mu(\Om)}\sum_{x\in H^k(\Om,r;G)}f(\rho(x))\eq
and for the inner product
\bq \li f,g\ri=\mu(\Si)\sum_{x\in H^k(\Si,r;G)}\ovl{f(x)}g(x).\eq
Here
\bq \mu(\Om)={|H^{k-1}(\Om,r;G)||H^{k-3}(\Om,r;G)|\dots\over
|H^{k-2}(\Om,r;G)||H^{k-4}(\Om,r;G)|\dots}\eq
(and the the same for $\Si$). Perhaps this $\mu$ is not a number you would like 
to meet in a dark forest, but this should not hide the simplicity of the thing. 
The gluing property follows from the exact sequence for the triple 
$r_{glued}\ss\Om\cup r_{glued}\ss\Om_{glued}$ ($r_{glued}$ is the red part of 
$\Om_{glued}$; $\Om\ss\Om_{glued}$ is achieved by separating slightly the glued 
yellow surfaces). Of course, the expression for $\mu$ was actually derived from 
this sequence.

This TFT reformulation of KW duality will be our starting point for non-abelian generalizations. Let us first have a look to see if we can recover the numbers $k$ and $l$ and the group $G$ from the TFT. It is enough to take yellow $(k+l)$-dim balls 
as $\Si$'s. The ball should be painted as follows: let us choose integers $k',l'$ such that $k'+l'=k+l$; we take a 
$S^{k'-1}\ss\partial\Si$ and paint its tubular neighbourhood in $\partial\Si$ in 
red; the rest (a tubular neighbourhood of a $S^{l'-1}$) is in black. Let us 
denote this $\Si$ as $\Si_{k',l'}$. The corresponding Hilbert space is trivial 
(equal to $\Bbb C$) if $k'\ne k$; if $k'=k$, it is the space of functions on $G$. 
The reader may try to define the Hopf algebra structure on this space using 
pictures (the $(1,1)$-case is drawn in the next section).

Our TFTs are of a rather special nature, because of the excision property of relative cohomology. It gives rise to the {\it squeezing property} of our TFTs. It is 
best explained by using an example. Imagine this full cylinder (the upper half of its mantle is red and the lower half is black; the invisible base is yellow):

$$\epsfxsize 5.5cm \epsfbox{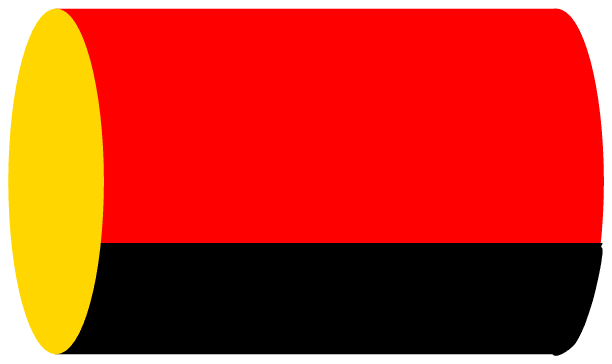} $$
We shall squeeze it in the middle, putting one finger on the red top and the 
other on the black bottom. The result is no longer a manifold---it has a 
rectangle in the middle (red from the top and black from the bottom), but it is 
surely homotopically equivalent (as a pair $(\Om,r)$, or as a pair $(\Om,b)$) to the original cylinder. Since we use relative cohomologies, the 
rectangle may be removed (it does not matter whether the cohomologies are 
relative to $r$ or to $b$ (the dual picture), as the rectangle is 
both red and black). The result is again a manifold of the type we admit:

$$\epsfxsize 5.5cm \epsfbox{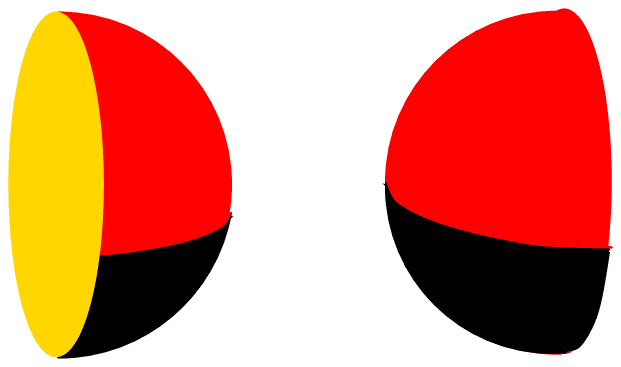}$$

Or, as another example: if our fingers are not big enough, we do not separate the cylinder into two 
parts, but instead we produce a hole in the middle (the top view of the result 
would be a red stain with a hole in the middle).

A bit informally the squeezing 
property can be formulated as follows: if a (hyper)surface appears as a result of squeezing $\Om$, red from one side and   black  from the other side, it 
may be removed.

Those TFTs that satisfy the squeezing property may be considered as generalizations of relative cohomology and of KW duality. As we shall see in the next setion, in the $(1,1)$-case they yield the expected result. Here is an example of such a TFT that does not come from an abelian group. We take 
a finite group $G$ and  two subgroups $R,B\ss G$ such that 
$RB=G$, $R\cap B=1$.  We shall consider principal $G$-bundles 
with reduction to $R$ over $r$ and to $B$ over $b$. If $P$ is such a thing, let 
$\mu(P)$ be the number of automorphisms of $P$. If $M$ is a space with some red 
and some black parts, let $P(M)$ be the set of isomorphism classes of these 
things. We 
set $\Hi(\Si)$ (the Hilbert space) to be the space of functions on $P(\Si)$ with 
the inner product \bq\li f,g\ri=\sum_{P\in P(\Si)}\mu(P)\overline{f(P)}g(P)\eq 
and finally, if $f\in\Hi(y)$, we set
\bq Z_\Om(f)=\sum_{P\in P(\Om)}{1\over\mu(P)}f(P|_y).\eq
This is surely a TFT. The squeezing property holds, because if we have a 
reduction for both $R$ and $B$ (as we have on the surfaces that appear by 
squeezing), these two reductions intersect in a section of the $G$-bundle. If 
$R=1$ and $B=G$, this TFT describes interacting $G$-spins (as in the 
introduction); the general case is more interesting, and we will meet its version in \S4.

\section{Non-abelian $(1,1)$-duality}

There are classical models (those appearing in Poisson--Lie T-duality \cite{PL}) 
that suggest a non-abelian generalization of $(1,1)$ KW duality. The PL T-duality 
generalizes the usual $R\leftrightarrow1/R$ T-duality, replacing the two circles 
(or tori) by a pair of mutually dual PL groups. Clearly, we have to replace the 
pair $G$, $\ti G$ by a pair of mutually dual quantum groups. This is not an easy 
(or well-defined) task. We have to define and to {\it understand} 
cohomologies with quantum coefficients. 

Here is how pictures solve this problem in a very simple way: just take a TFT in three dimensions, 
satisfying the squeezing property. A finite quantum group (finite-dimensional Hopf $C^*$-algebra) will appear 
independently of the classical motivation. If you exchange red and black (which 
gives a new TFT), the quantum group will be replaced by its dual. This is the 
non-abelian (or quantum) $(1,1)$ KW duality.

Now we will draw the pictures. I learned this 3d way of representing quantum 
groups at a lecture by Kontsevich \cite{Kon}; it was one of the sources of this 
work. The finite quantum group itself is $\Hi(\Si_{1,1})$. The product 
$\Hi(\Si_{1,1})\otimes\Hi(\Si_{1,1})\map\Hi(\Si_{1,1})$ is on this picture:

$$\epsfxsize 5.5cm \epsfbox{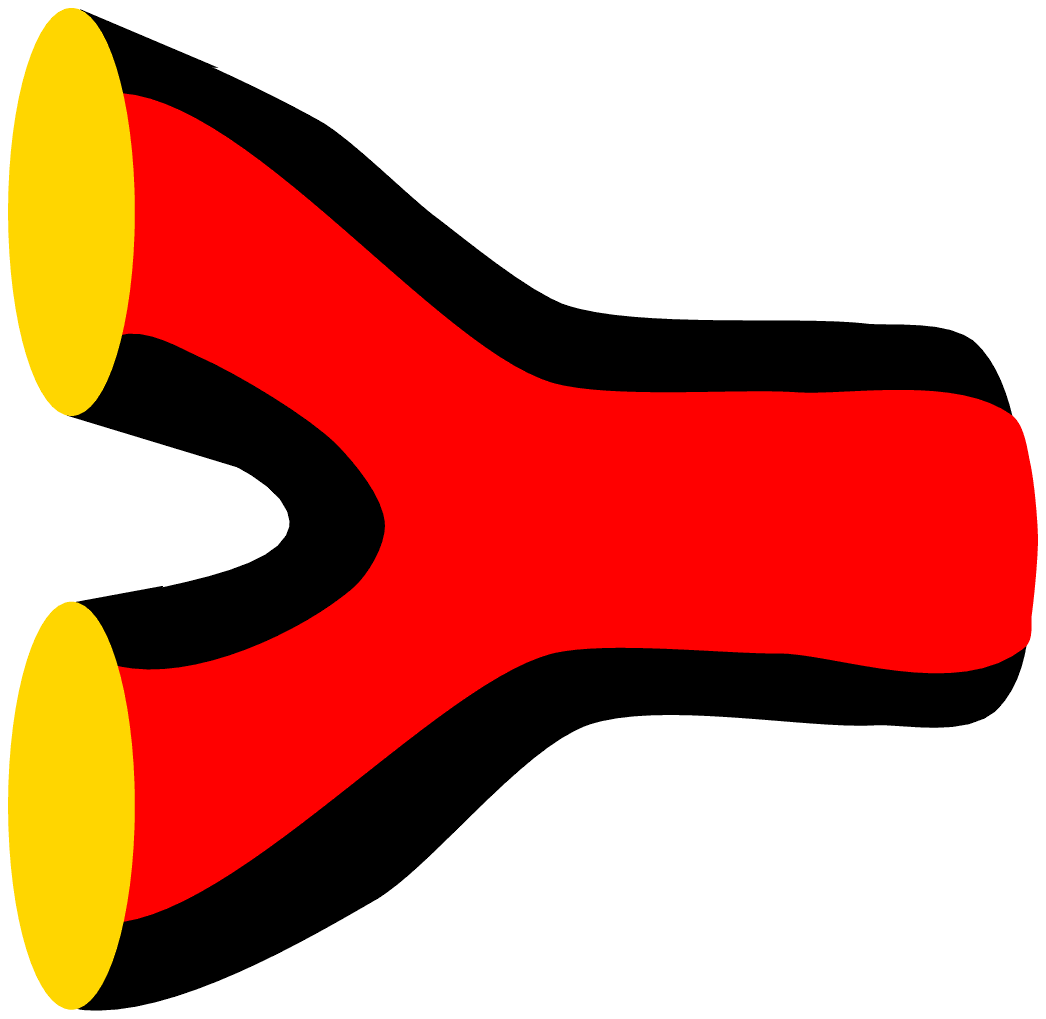}$$

And here are all the operations. Coloured 3d objects are hard to draw (but not 
hard to visualize!); imagine that the pictures represent balls and that their 
invisible sides are completely yellow. The antipode $S$ is simply the half-turn, the 
involution $*$ is the reflection with respect to the horizontal diameter, and 
the rest is on the figure:

$$\epsfxsize 5cm \epsfbox{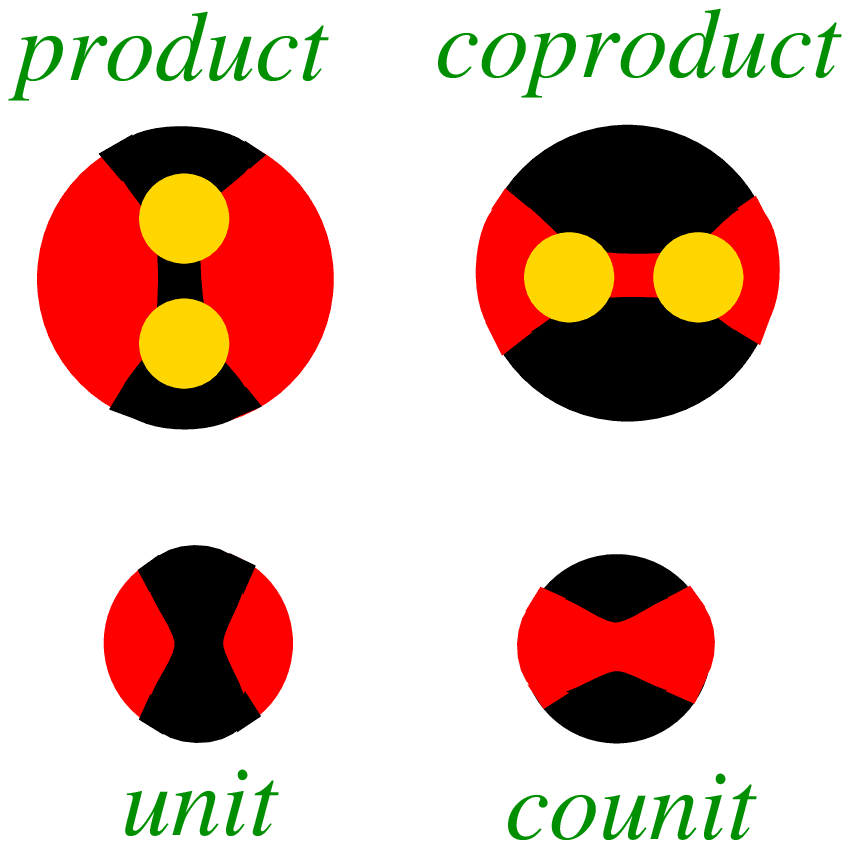}$$

Why is it a quantum group? Just imagine the pictures representing the axioms and 
use the squeezing property in a very simple manner.

Let us make a conjecture that there is a 1--1 correspondence between finite 
quantum groups and 3d TFTs satisfying the squeezing property, with trivial 
(i.e. one-dimensional) $\Hi(\Si_{0,2})$ and $\Hi(\Si_{2,0})$. To support the conjecture, finite quantum groups are in 1--1 correspondence with modular functors of a certain kind (cf. \cite{KW}), clearly connected with our TFTs.

\section{Chern--Simons with coloured boundary}

Let us recall a basic analogy between symplectic manifolds and vector spaces (the aim of quantization is to go beyond a mere analogy):

\begin{center}
\begin{tabular}{|c|c|} \hline
{\em Vector}&{\em Symplectic}\\ \hline\hline
Vector space&Symplectic manifold\\ \hline
Vector&Lagrangian submanifold\\ \hline
$V_1\otimes V_2$&$M_1\times M_2$\\ \hline &\\[-2.5ex]
$V^*$&$\overline{M}$\\ \hline
Composition of linear maps&Composition of Lagrangian relations\\ \hline
\end{tabular}
\end{center}

\def\fg{{\frak g}}
\def\fb{{\frak b}}
\def\fr{{\frak r}}
One can easily describe the symplectic analogue of the Chern--Simons TFT (see e.g. \cite{Fr}). Let $\fg$ be a Lie algebra with invariant inner product. If $\Si$ is a closed oriented surface then the moduli space of flat $\fg$-connections is a symplectic manifold (with singularities). The symplectic form is given as follows. The vector space
of all $\fg$-valued 1-forms on $\Si$ is symplectic, with the symplectic form 
\bq\om(\al_1,\al_2)=\int_\Si\langle\al_1,\al_2\rangle.\eq
When we restrict ourselves to flat connections, the space is no longer symplectic, but the null directions of the 2-form give just the orbits of the gauge group, so the quotient (the moduli space) is symplectic. Let us denote it by $M_\Si$.

We have associated a symplectic space to every oriented closed surface. Now, if $\Om$ is an oriented compact 3-fold with boundary $\Si$, we should find a Lagrangian subspace $\La_\Om\ss M_\Si$. Indeed, $\La_\Om$ consists just of those flat connections on $\Si$, which can be extended to $\Om$.

Let us make a minute extension of this construction, allowing a boundary coloured in red and black. Let $\fb,\fr\ss\fg$ be a Manin triple. We shall consider flat $\fg$ connections as before, with the obvious boundary conditions---on the red part of the boundary the connection should take values in $\fr$ and on the black part in $\fb$. Similarly, the gauge group consists of the maps to $G$ with the same boundary conditions. This really defines a symplectic TFT for our pictures. From this symplectic TFT we obtain a symplectic analogue of quantum group (using the pictures of the previous section). One readily checks that it is the double symplectic groupoid of Lu and Weinstein \cite{LW}---the symplectic analogue of the quantum group coming from the Manin triple $\fb,\fr\ss\fg$.

For this reason, it is reasonable to conjecture that perturbative quantization of our Chern--Simons TFT with boundary will give the corresponding quantum group.

In the next section we  return to the vector side of our table, to general 3d TFTs that satisfy the squeezing property. We shall see this connection with CS TFTs again, in a different guise.

\section{Pictures of the Drinfeld double}

There are lots of algebras, modules,  etc., in our pictures. We shall describe 
only the Drinfeld double, since it is important in PL T-duality, and also 
to make a connection with Reshetikhin--Turaev invariants. Here are the unit and the 
counit:

$$\epsfxsize 6cm \epsfbox{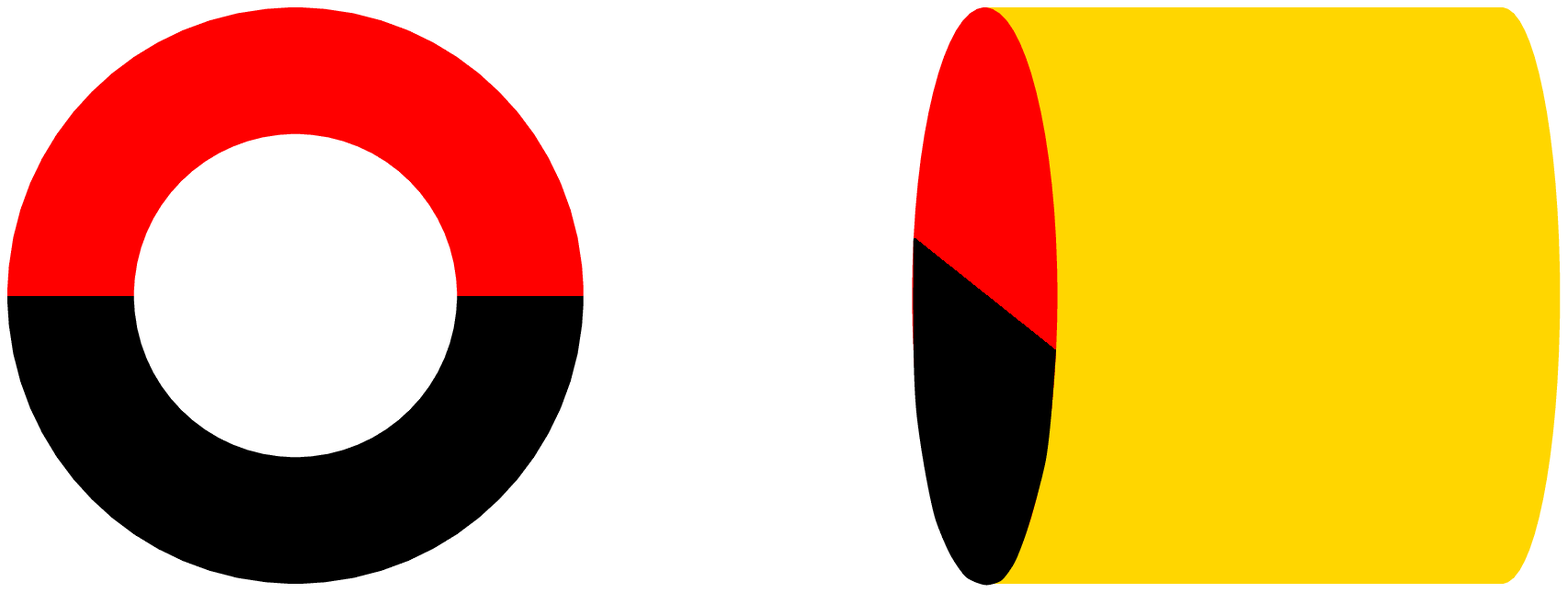}$$
The invisible side of the full torus on the first picture is yellow; this closed 
yellow strip is the double. On the second picture it is represented as the 
mantle of the cylinder (the invisible base of the cylinder is painted as the visible one).

Here is the product (the picture is yellow from the invisible side):
$$\epsfxsize 4cm \epsfbox{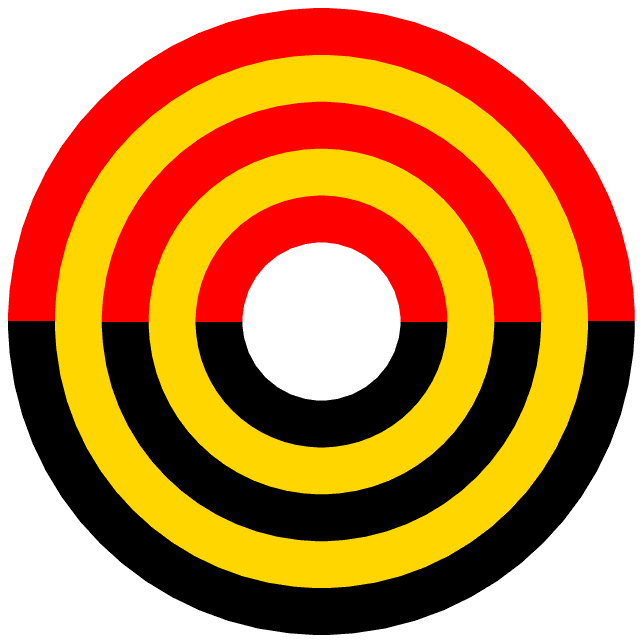}$$
and finally the coproduct:
$$\epsfxsize 4cm \epsfbox{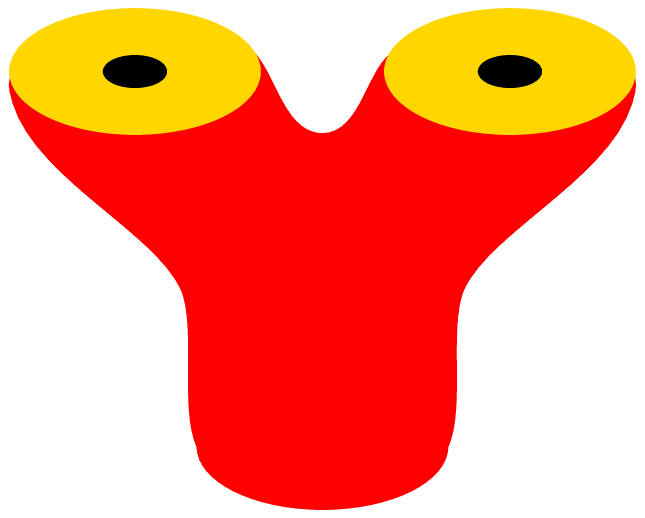}$$
This picture requires an explanation. It represents a thick Y from which a thin 
Y was removed (you can see it as the black holes in the yellow disks). The 
fronts of these Y's are red and their backs are black (the invisible bottom of 
the picture is yellow---it is the third double).

For completeness, the antipode is a half-turn and the involution a reflection, 
both exchanging the boundary circles of the double.

\font\cyr=wncyr10
Now we know the double as a Hopf algebra, but its real treasure is the 
$R$-matrix:
$$\epsfxsize 4cm \epsfbox{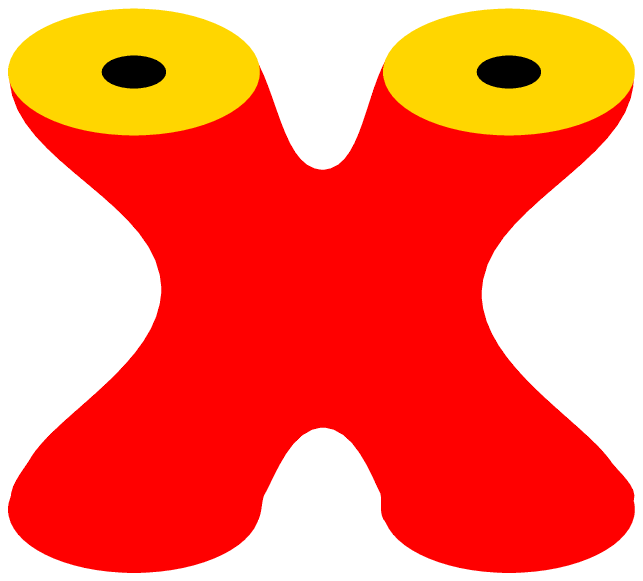}$$
It is quite similar to the Y-picture, but this time we do not remove a thin X, 
but rather two tubes connecting the top holes with the bottom ones. However, if 
one tube connected the left holes and the other one the right holes, the picture 
would not be very interesting. We could squeeze the X in the middle, dividing it 
into two vertical cylinders. We would simply have an identity. However, in the X of the $R$-matrix, the tubes are diagonal. There are two ways for them to avoid each 
other; one gives the $R$-matrix and the other its inverse.
This X has two incoming and two outgoing doubles; you can also imagine $n$ 
doubles at the bottom, tubes forming a braid inside and leaving the body at the 
top, in the middle of $n$ other doubles (the Cyrillic letter {\cyr ZH} is good here). We  directly see a representation of the braid group.

With this picture in mind, we can find the Reshetikhin--Turaev (RT) invariants coming from the double. Namely, the boundary-free part of our TFT is the Chern--Simons theory coming from the double. Here is a sketch of the proof: suppose $\Om$ is a closed oriented 3-fold with a ribbon link. We colour each  of the ribbons in red on one side and in black on the other side, blow it a little, so that the ribbon becomes a full torus removed from $\Om$, and paint on the torus a little yellow belt. Our TFT gives us an element of ${\rm double}^{\otimes n}$ (one double for each yellow belt), where $n$ is the number of components of the link. Actually, this element is from $(\mbox{centre of double})^{\otimes n}$ (we can move a yellow belt along the torus and come back from the other side). It is equal to the RT invariant. This claim follows immediately from the definition of RT invariants: If $\Om=S^3$, we are back in our picture of braid group, and generally, surgery along tori in $S^3$ can be replaced by gluing tori along the yellow belts.

Finally, we can get rid of red and black and instead consider $\Om$'s with boundary consisting of yellow tori: one easily sees that $\Hi(\mbox{yellow torus})= \mbox{centre of double}$.

\section{Conclusion: Higher dimensions?}

There are several open problems remaining. Apart from the mentioned conjectures there is a problem with the square of the antipode: for the naive definition of TFT used in this paper, it has to be 1. One should find a less naive definition and prove in some form the claim that our pictures are equivalent to Hopf algebras.

However, in spite of these open problems, the presented picture is very simple and quite appealing. It is really tempting 
(and almost surely incorrect) to suggest \bq\mbox{\it duality = TFT with the 
squeezing property.}\eq It would be nice to understand the basic building blocks 
of these TFTs that replace quantum groups in higher dimensions. It is a purely 
topological problem. It would also be nice to have a non-trivial example with 
non-trivial $\Hi(\Si_{2,2})$, to see an instance of S-duality ($(2,2)$-duality) in this way.

The field of duality is vast and connections with this work may be of diverse 
nature. But let us finish with a rather internal question: Why yellow, red 
and black?


\begin{thebibliography}{9}
\bibitem{Ati}M.F. Atiyah, {\it Topological quantum field theories}, Publ. Math. 
IHES {\bf 69} (1988) 175.
\bibitem{Fr}D.S. Freed, {\it Classical Chern-Simons theory I}, Adv. Math. {\bf 113} (1995) 237.
\bibitem{PL}C. Klim\v c\'\i k, P. \v Severa, {\it Dual non-Abelian duality and 
the Drinfeld double}, Phys. Lett. {\bf B351} (1995) 455.
\bibitem{Kon}M. Kontsevich, {\it Geometry of formulae}, lecture given at {\it 
Colloque scientifique du 40${}^{\,e}$ anniversaire}, IHES, 1998, 8 Oct.
\bibitem{LW} J.-H. Lu, A. Weinstein, {\it Groupo\"\i des symplectiques doubles des groupes de Lie-Poisson}, C. R. Acad. Sci. Paris, S\'er. I Math. {\bf 309} (1989), No. 18.
\bibitem{RT}N.Y. Reshetikhin, V.G. Turaev, {\it Invariants of three-manifolds 
via link polynomials and quantum groups}, Invent. Math. {\bf 103} (1991) 547.
\bibitem{KW}P. \v Severa, {\it Quantum Kramers--Wannier duality and its 
topology}, hep-th/9803201.
\bibitem{Wit}E. Witten, {\it Quantum field theory and the Jones polynomial}, 
Commun. Math. Phys. {\bf 121} (1989) 351.
\end{thebibliography}
\end{document}